\documentclass[preprint]{aastex}
\usepackage{natbib}
\usepackage[american]{babel}
\usepackage{subfigure}
\usepackage{amsmath}
\usepackage{amssymb}
\usepackage{graphicx}

\shorttitle{H.E.S.S. observations of the globular clusters NGC~6388 and M~15 
       and search for a Dark Matter signal}
\shortauthors{H.E.S.S. collaboration: A.~Abramowski et al.}

\begin{document}
\bibliographystyle{natbib}
\title{H.E.S.S. observations of the globular clusters NGC~6388 and M~15 \\ 
       and search for a Dark Matter signal}

\author{HESS Collaboration:
A.~Abramowski\altaffilmark{1},
F.~Acero\altaffilmark{2},
F.~Aharonian\altaffilmark{3,4,5},
 A.G.~Akhperjanian\altaffilmark{6,5},
 G.~Anton\altaffilmark{7},
 A.~Balzer\altaffilmark{7},
 A.~Barnacka\altaffilmark{8,9},
 U.~Barres de Almeida\altaffilmark{10,\ast},
 A.R.~Bazer-Bachi\altaffilmark{11},
 Y.~Becherini\altaffilmark{12,13}
% ...
}

 \altaffiltext{1}{Universit\"at Hamburg, Institut f\"ur Experimentalphysik, Luruper Chaussee 149, D 22761 Hamburg, Germany}
 \altaffiltext{2}{Laboratoire de Physique Th\'eorique et Astroparticules, Universit\'e Montpellier 2, CNRS/IN2P3, CC 70, Place Eug\`ene Bataillon, F-34095 Montpellier Cedex 5, France} 
 \altaffiltext{3}{Max-Planck-Institut f\"ur Kernphysik, P.O. Box 103980, D 69029 Heidelberg, Germany}
\altaffiltext{4}{Dublin Institute for Advanced Studies, 31 Fitzwilliam Place, Dublin 2, Ireland}
\altaffiltext{5}{National Academy of Sciences of the Republic of Armenia, Yerevan}
\altaffiltext{6}{Yerevan Physics Institute, 2 Alikhanian Brothers St., 375036 Yerevan, Armenia} 
\altaffiltext{7}{Universit\"at Erlangen-N\"urnberg, Physikalisches Institut, Erwin-Rommel-Str. 1, D 91058 Erlangen, Germany }
\altaffiltext{8}{Nicolaus Copernicus Astronomical Center, ul. Bartycka 18, 00-716 Warsaw, Poland} 
\altaffiltext{9}{IRFU/DSM/CEA, CE Saclay, F-91191 Gif-sur-Yvette, Cedex, France}
\altaffiltext{10}{University of Durham, Department of Physics, South Road, Durham DH1 3LE, U.K.}
\altaffiltext{11}{Centre d'Etude Spatiale des Rayonnements, CNRS/UPS, 9 av. du Colonel Roche, BP 4346, F-31029 Toulouse Cedex 4, France}
\altaffiltext{12}{Astroparticule et Cosmologie (APC), CNRS, Universit\'e Paris 7 Denis Diderot, 10,
rue Alice Domon et Leonie Duquet, F-75205 Paris Cedex 13, France.
 Also at UMR 7164 (CNRS, Universit\'e Paris VII, CEA, Observatoire de Paris)}
\altaffiltext{13}{Laboratoire Leprince-Ringuet, Ecole Polytechnique, CNRS/IN2P3,
 F-91128 Palaiseau, France}
% ...
\altaffiltext{$\ast$}{supported by CAPES Foundation, Ministry of Education of Brazil}
% ...

\begin{abstract}
Observations of the globular clusters NGC~6388 and M~15 were carried out by the H.E.S.S. array of Cherenkov telescopes for a live time of 27.2 and 15.2 hours respectively. No gamma-ray signal is found at the nominal target position of NGC~6388 and M~15.  
In the primordial formation scenario, globular clusters are formed in a dark matter halo and 
dark matter could still be present in the baryon-dominated environment of globular clusters. 
This opens the possibility of observing a dark matter self-annihilation signal. 
The dark matter content of the globular clusters NGC~6388 and M~15
is modelled taking into account the astrophysical processes that can be expected to influence the dark matter distribution during the evolution of the 
globular cluster: the adiabatic contraction of dark matter by baryons, the adiabatic growth of a black hole in the dark matter halo and the kinetic heating of dark matter by stars.  95\% confidence level exclusion limits on the dark matter particle velocity-weighted annihilation cross section are derived for these dark matter haloes.
In the TeV range, the limits on the velocity-weighted annihilation cross section are derived at the $\rm 10^{-25}\,cm^{3}s^{-1}$ level and a few $\rm 10^{-24}\,cm^{3}s^{-1}$ for NGC~6388 and M~15 respectively. 
\end{abstract}


\section{\label{sec:introduction}Introduction}
Dark matter (DM) is expected to play a key role in the dynamics of a wide range of systems, from galactic object scale to galaxy cluster scale \citep{Bertone:2004pz}. 
The inner parts of galaxies were used extensively to search for DM even though they are not necessarily dominated by DM \citep{Englmaier:2006pa}. 
Observations by the H.E.S.S. experiment  towards the Galactic Center reveal  
a very high energy (VHE, $\rm E_{\gamma}\gtrsim$ 100~GeV) gamma-ray signal with the flux level 
$\rm \Phi(>1TeV) \simeq 2\times 10^{-12}\,cm^{-2}s^{-1}$ \citep{Aharonian:2009zk}.
The plausible DM-originated fraction of this signal
is entirely dominated by standard astrophysical emission processes \citep{Aharonian:2006wh}. 
As opposed to this, dwarf galaxies are believed to be among the most DM-dominated objects and are believed to have a reduced astrophysical background \citep{Mateo:1998wg}
since these objects have little or no recent star formation activity.
Nearby dwarf galaxies in the Local Group have already been observed with H.E.S.S., {\it i.e.} Sagittarius, Canis Major, Sculptor and Carina \citep{Aharonian:2007km, Aharonian:2008dm, Aharonian:2010zzt}, yielding no signal. Exclusion limits on the velocity-weighed annihilation cross section between $\sim$10$^{-24}$ to 
$\sim$10$^{-22}$\,cm$^3$s$^{-1}$ have been reported in the TeV range.

Several Galactic Globular Clusters (GCs) have been observed with ground-based Cherenkov telescopes and upper limits on $\gamma$-ray emission from standard astrophysical processes 
have been reported on Omega Centauri, 47 Tucan\ae, M~13, M~15 and M~5 \citep{Kabuki:2007am,Aharonian:2009nh,Anderhub:2009px,McCutcheon:2009hs}.
GCs are also potential targets for indirect DM searches (\citeauthor{Wood:2008hx} \citeyear{Wood:2008hx}). 
They are dense stellar systems of $\gtrsim$~10~Gyr old, found 
in haloes of galaxies, with typical masses between 10$^4$ and a few 10$^6$~M$_{\odot}$, similar to dwarf galaxies.
However, GCs are much more compact than dwarf galaxies. 
Observations of GCs do not suggest the presence of a significant amount of DM,
but rather that these objects are dominated by baryons \citep{binneytremaine}.
In the primordial formation scenario of GCs \citep{Peebles:1984zz}, GCs were formed 
in DM minihaloes before or during the reionization \citep{Komatsu:2008hk},
before formation of galaxies. However the distribution of GC colors \citep{Brodie:2006zz} 
suggests that only metal poor clusters have a cosmological origin while metal
rich clusters formed in star-forming events such as galaxy-galaxy mergers. 
The existence of an extended dark halo required by the GC primordial formation scenario, 
has been challenged recently  by \citeauthor{Baumgardt:2009hv} (\citeyear{Baumgardt:2009hv})
and \citeauthor{Conroy:2010bs} (\citeyear{Conroy:2010bs}). 
They show that  the stellar kinematics of NGC 2419, a remote GC which experiences little tidal effects from the Milky Way,
is incompatible with the presence of an extended dark halo. However,  \citeauthor{Cohen:2010gq} (\citeyear{Cohen:2010gq})
 have shown that the measured spread in the Ca abundance can be only explained if NGC 2419 is the remnant of  a more massive object.

In the purpose of the paper, GCs are assumed to have formed in DM minihaloes and thus were DM-dominated in their primordial stage.
Note that M~15 is a metal-poor GC, [Fe/H]$\simeq$-2.37 \citep{Harris:1996},  while NGC~6388 is metal-rich, [Fe/H]$\simeq$-0.55 \citep{Harris:1996}, 
so the DM minihalo scenario is better motivated for M~15 than NGC~6388. However, NGC~6388 might host a 
$\gtrsim$ 10$^3$\,M$_{\odot}$  black hole (\citeauthor{Lanzoni:2007ms} \citeyear{Lanzoni:2007ms}). Such
 massive ($\gtrsim$ 10$^3$ M$_{\odot}$) black holes are not easily formed 
in star-forming events \citep{Yungelson:2007qu} suggesting a primordial formation origin.
During the evolution of the globular cluster, the DM reacts to the infall of baryons and is pulled in
towards the center. This process is usually  referred to as the adiabatic contraction (AC) model \citep{Blumenthal:1985qy, Zeldovich:1980st, Jesseit:2002tj, Gnedin:2004cx,Prada:2004pi}. 
The effect of the contraction of DM in response to the baryon infall is
particularly important for the calculation of the DM annihilation in baryonic environments
such as the Galactic Center region \citep{Gnedin:2003rj,Prada:2004pi}.

The distribution of baryons and DM is affected by the kinetic heating of DM by baryons (\citeauthor{Merritt:2003qk}  \citeyear{Merritt:2003qk}) and by the presence of 
a black hole (BH) \citep{Gondolo:1999ef}. 
A growing body of observations on GCs shows that they may harbor intermediate mass black holes (IMBHs) 
with masses ranging from 10$^3$ to 10$^5$~M$_{\odot}$,  although the existence of these objects is not yet established.
Indirect evidence includes the extrapolation of the empirical relation M$_{\rm BH}$-M$_{\rm Bulge}$ found for the supermassive BHs in galactic nuclei, 
which leads naturally to the prediction of existence of IMBHs \citep{Magorrian:1997hw}. 
Besides, ultra-luminous X-ray sources (ULXs) apparently not associated to active galactic nuclei 
are proposed to be related to IMBHs \citep{Colbert:2002mi,Swartz:2004xt,Dewangan:2005kj}. 
From the theoretical side, the existence of IMBHs is a generic assumption of scenarios that seek to explain the formation of supermassive BHs from the accretion and merging of massive seeds \citep{islama:2003,volonteri:2003,Koushiappas:2003zn}. 
Several globular clusters may host IMBHs: 
NGC~6388 (\citeauthor{Lanzoni:2007ms} \citeyear{Lanzoni:2007ms}), $\omega$~Centauri \citep{Noyola:2008kt} in the Milky Way or even G1 in M~31 \citep{Kong:2007mu,Ulvestad:2007cw}.

The paper is structured as follows. Section 2 describes the observations carried out with the H.E.S.S. experiment and 
the analysis of the data on the two Galactic GCs NGC~6388 and M~15. 
In section 3, the DM halo modelling relevant for the two GCs is briefly presented 
(more details are given in the appendix)
and exclusion limits on the velocity-weighted annihilation cross section of DM particles are derived for realistic DM halo profiles. 
Finally, the results are summarized in section 4.

\section{\label{sec:datanalysis}Observations and data analysis}
The H.E.S.S. (High Energy Stereoscopic System) array of Cherenkov telescopes is located in the Khomas Highland of Namibia at an altitude of 1800 m above sea level. The system consists of four Imaging Atmospheric Cherenkov telescopes of 12~m diameter ($\sim$107\,m$^2$) each. 
The total field of view of H.E.S.S. is 5$^{\circ}$ in diameter.
The H.E.S.S.  instrument achieves an angular resolution of 5' per gamma-ray and a  point-like source sensitivity at the level of $\rm 10^{-13}\,cm^{-2}s^{-1}$ above 1 TeV for a 5$\sigma$ detection in 25 hours at an observation zenith angle of 20$^{\circ}$ \citep{deNaurois:2009ud}. 

For both M~15 and NGC~6388, data are taken in {\it wobble mode} \citep{Aharonian:2006pe},
where the pointing direction is chosen at an alternating offset of $\pm$0.7$^{\circ}$ to $\pm$1.0$^{\circ}$ from the target position.
Standard quality selection \citep{Aharonian:2006pe} is applied to the data.
After the calibration of the raw data from PMT signals, the events are reconstructed using a technique based 
on a semi-analytical shower development model \citep{deNaurois:2009ud}. This reconstruction method yields a relative energy resolution of $\sim$10\% 
and an angular resolution  of 0.06$^{\circ}$ per gamma-ray (68\% containment radius).
The background level is estimated using the ring-background method \citep{Puhlhofer:2003ir, Berge:2006ae}.
 The core of the GCs is very small in comparison to the H.E.S.S. point spread function. 
They would thus be seen as point sources by H.E.S.S..
The source region referred to hereafter as  the ON region, is defined as a circular region of 0.07$^{\circ}$radius around the target position. 
The background region is defined as an annulus around the pointing position with a medium radius equal to the distance of the target and a half-thickness of 0.07$^{\circ}$. 
It is referred to hereafter as  the OFF region.
The cut values used to select the gamma-ray events are shown in Tab.~\ref{tabcut}.
 Given the Galactic latitude of the two Galactic GCs, contamination by diffuse TeV gamma-ray emission is very unlikely.
The results presented below have been cross-checked by an independent analysis chain \citep{Berge:2006ae,Aharonian:2004ap}. Both analyses give consistent results.

\subsection{NGC~6388}
NGC~6388 is one of the best known Galactic GC. It is located at $\sim$\,11.5\,kpc from the Sun,  
at RA = $\rm 17^h36^m17.05^s$ and Dec = $\rm -44^{\circ}44'05.8''$ (J2000),  and has a mass
estimated to be $\sim$\,10$^6$~M$_{\odot}$. 
The stellar mass density in the core, with radius $r_{\rm c}$ = 0.4\,pc (0.12 arcminutes), reaches $\sim$\,5$\times$10$^5$\,M$_{\odot}$pc$^{-3}$. The tidal radius 
is $r_{\rm t}\,\sim$\,25\,pc (\citeauthor{Lanzoni:2007ms} \citeyear{Lanzoni:2007ms}). 
The structural properties of NGC~6388 are summarized in Tab.~\ref{tab0}.
Using the high-resolution HST and WFI observations at ESO, \citeauthor{Lanzoni:2007ms}(\citeyear{Lanzoni:2007ms})
show that the surface brightness density of stars significantly deviates from a flat core in the inner part, which is
compatible with the existence of an IMBH with a mass of  $\sim\rm 5\times10^3\,M_{\odot}$. 
A power law with a slope of -0.2 is detected in the surface brightness density profile, 
which suggests the presence of a central IMBH \citep{Baumgardt:2004tp,Noyola:2006mg,Umbreit:2008nm}.
The \emph{Chandra} satellite has detected three X-ray sources, coincident in position with the centre of gravity of NGC~6388 
located with an uncertainty of $\rm 0.3''$.  
One of these may be the X-ray counterpart of the putative IMBH (\citeauthor{Nucita:2007yn}\citeyear{Nucita:2007yn}).

\begin{figure}[]
\begin{center}
\epsscale{1.}
\subfigure[NGC6388]{\includegraphics[width=8cm]{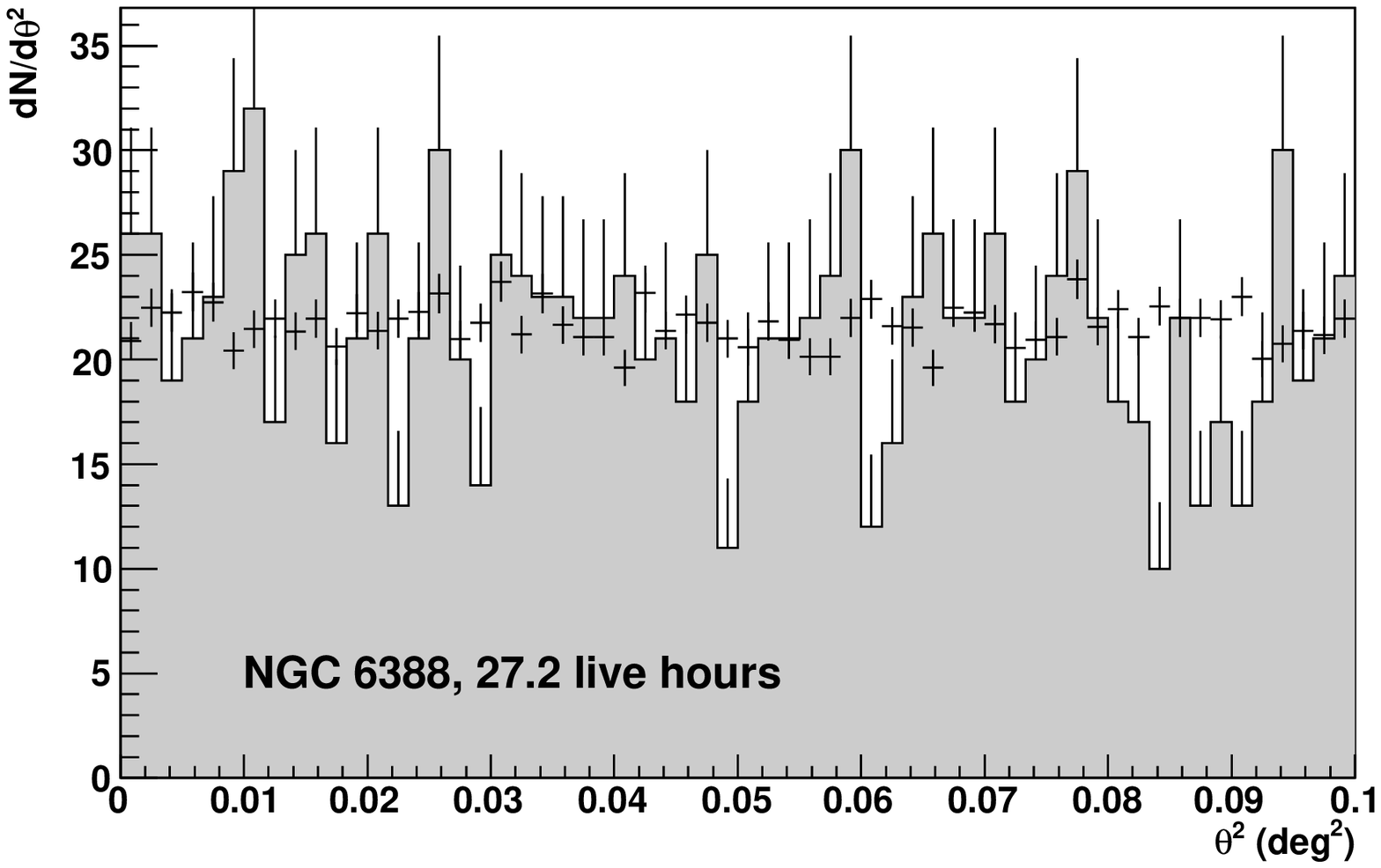}}
 \subfigure[M15]{\includegraphics[width=8cm]{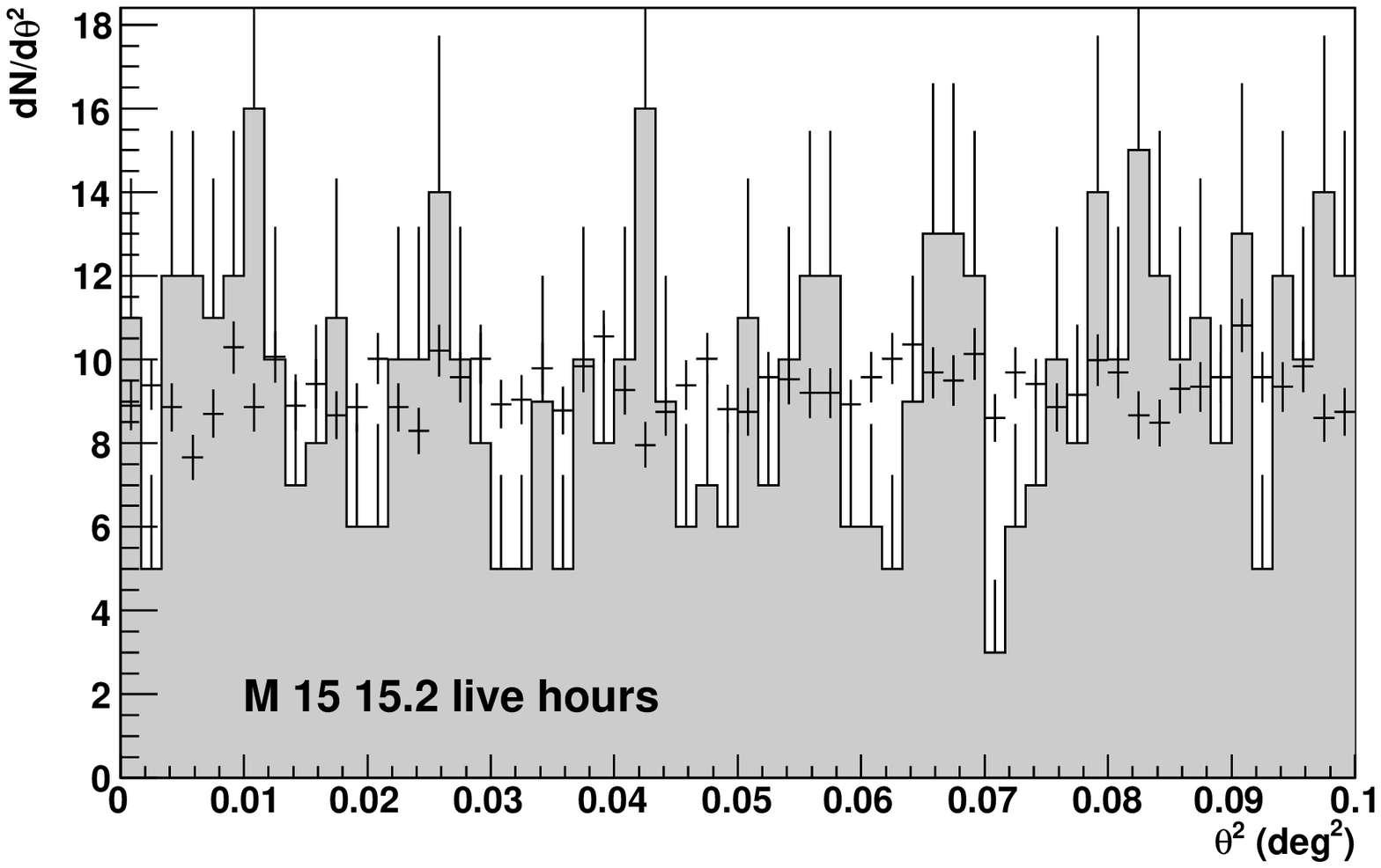}}
\end{center}
\caption{
$\theta^2$ radial distribution of the ON (gray histogram) and normalized OFF (black crosses) events for gamma-ray like events from a
0.07$^{\circ}$ radius region around the  target position
for (a) NGC~6388 (RA=$\rm17^h36^m17.05^s$ and Dec=$\rm -44^{\circ}44'05.8''$, J2000) and (b) M~15 ($\rm RA=21^h28^m58.3^s$ 
and $\rm Dec=12^{\circ}10'00.6''$, J2000). The ON regions correspond to a maximum $\theta^2$ of 0.005 deg$^{2}$.
No significant excess is found in the ON region for NGC~6388 or M~15. \label{fig:Theta2plot}}
\end{figure}

The H.E.S.S. observations of NGC~6388 were taken between June 2008 and July 2009. 
The observation zenith angles range from 20$^{\circ}$ to 44$^{\circ}$ with
a mean zenith angle of 22.9$^{\circ}$, and  the exposure time is 27.2 hours.  
The left-hand side of Fig.~\ref{fig:Theta2plot} shows the $\theta^2$ radial distribution of the ON and OFF events for NGC~6388.
N$_{\rm ON}$ = 71 gamma-ray candidates were found in the ON region 
while the measured number of OFF events is N$_{\rm OFF}$ = 1754. 
With the ratio of the off-source solid angle region to the on-source solid angle region $\alpha$ = 26.5, the expected number of background events in the ON region is 66.1.
Since N$_{\rm ON}$ and N$_{\rm OFF}$/$\alpha$ are compatible at the 2$\sigma$ level, no significant gamma-ray excess is found above the background. 
An upper limit at 95\% confidence level (C.L.) on the number of gamma-rays can be derived  using the method 
developed in \citeauthor{rol05}(\citeyear{rol05}): $\rm N_{\gamma}^{95\% C.L.}$ = 21.6.

\subsection{M~15}
M~15 (NGC~7078) is a well-studied Galactic GC centered at the position RA = $\rm 21^h28^m58.3^s$ 
and Dec = $\rm 12^{\circ}10'00.6''$ (J2000). It is situated at $\sim$\,10\,kpc from the Sun. Its estimated mass is $\sim$\,5$\times$10$^5$\,M$_{\odot}$. 
The stellar mass density in the core with radius $r_{\rm c}$ = 0.04 pc is about~10$^7$\,M$_{\odot}$pc$^{-3}$ (\citeauthor{1997ApJ...481..267D}\citeyear{1997ApJ...481..267D}), and the tidal radius is $r_{\rm t}\,\sim$\,30\,pc. 
The structural properties of M~15 are summarized in Tab.~\ref{tab0}.
The surface brightness density of the GC M~15 suggests the presence of a stellar cusp in the inner part, 
at least down to distances of a few 10$^{-2}$\,pc (\citeauthor{1997ApJ...481..267D}\citeyear{1997ApJ...481..267D}).
M~15 may thus harbor an IMBH \citep{Gerssen:2002iq,Kiselev2008} in its center.
 However, the study on milli-second pulsars in M~15 sets an upper
limit of 10$^3$\,M$_{\odot}$ on the mass of a hypothetical central BH \citep{DePaolis:1996tq}. In what follows, no central back hole is assumed for the modelling of M~15.

The observations of M~15 by H.E.S.S. were carried out in 2006 and 2007 with an  offset angle of 0.7$^{\circ}$ and zenith angles from 34$^{\circ}$ to 44$^{\circ}$
resulting in 15.2 hours of high quality data at a mean zenith angle of 37.0$^{\circ}$. 
The right-hand side of Fig.~\ref{fig:Theta2plot} shows the $\theta^2$ radial distribution of the ON and OFF events for M~15.
The data analysis reveals no significant gamma-ray signal at the nominal position. With N$_{\rm ON}$ =  28, N$_{\rm OFF}$ = 719 and $\alpha$ = 26.6, the expected number of background events in the ON region is 27.0. The 95\% C.L. upper limit on the number of gamma-rays is $\rm N_{\gamma}^{95\% C.L.}$ = 11.5. 
Tab.~\ref{tab1} summarizes observations and upper limits on count numbers for NGC~6388 and M~15.

\section{\label{sec:DM}Dark matter constraints}
\subsection{Dark matter halo modelling} \label{sec:dmm}
The relaxation time,  defined in Eq.~(\ref{eq:relaxationtime}) of the
appendix, has a much smaller value, $T_{r}\sim$~10$^7$\,yr, in GCs than 
in galaxies, where it is typically of the order of 10$^{13}$~yr \citep{binneytremaine}. 
Since GCs are among the oldest objects known, their present DM density depends 
on their history and evolution. 
During infall events such as core collapses (\citeauthor{spitzer} \citeyear{spitzer}), the DM is compressed towards the center following the AC 
scenario \citep{Zeldovich:1980st,Blumenthal:1985qy}. This profile is referred to hereafter as the AC NFW profile. But the kinetic heating of DM particles by stars (\citeauthor{Merritt:2003qk}  \citeyear{Merritt:2003qk})  
tends to wash out the adiabatic contraction effect over a timescale of the order of T$_{\rm r}$. 
Both effects were taken into account in the modelling of M~15 and NGC~6388, 
following the approach of \citeauthor{Merritt:2006mt}  (\citeyear{Merritt:2006mt}) and \citeauthor{Bertone:2008a} (\citeyear{Bertone:2008a}).

As mentioned in the introduction, the primordial formation scenario
of globular clusters \citep{Peebles:1984zz} assumed here requires that
globular clusters were formed in extended DM haloes.
The DM halo profile of a globular cluster is thus  modelled assuming 
an initial Navarro-Frenk-White (NFW) profile (\citeauthor{Navarro:1996gj} \citeyear{Navarro:1996gj}) described by:
\begin{equation}
\label{eq:rhonfw}
 \rho(r) = \rho_{\rm 0}\left(\frac{r}{r_{\rm s}}\right)^{-1}\left(1+\frac{r}{r_{\rm s}}\right)^{-2}\, .
\end{equation}
This DM halo is parameterized by a virial mass\footnote{M$_{\rm vir}$ is defined as the mass inside the radius R$_{\rm vir}$ assuming a mean density equal to 200 times the critical density of the Universe \citep{pdg}.}
 $M_{\rm vir}$ and a concentration parameter $c_{\rm vir}$. The normalization parameter $\rho_{\rm 0}$ and the scale radius
$r_{\rm s}$ can be related to the virial mass and the concentration
parameter using the following relations (\citeauthor{Navarro:1996gj} \citeyear{Navarro:1996gj})
\begin{equation}
\label{eq:rho0} \rho_0 = \frac{M_{\rm vir}}{4\pi r_{\rm s}^3
f(c_{\rm vir})}\,\,\,\,\,,\,\,\,\,\,r_{\rm s} = \frac{R_{\rm vir}}{c_{\rm
vir}}\,\,; 
\end{equation}
where the function $f(x)$ is, neglecting constants, the volume integral of the NFW profile given by  
$f(x)  \equiv {\rm ln}(1+x)-x/(1+x)$. 
The present baryonic mass of the GC provides a lower bound on its virial mass.
Besides, for $M_{\rm vir}$ larger than 10$^8$~M$_{\odot}$  the GC would be expected
to spiral towards the center of the Milky Way in less than the age
of the Universe. Conservative values for $M_{\rm vir}$ lie therefore in the
range [$\rm 5\times 10^6 - 5\times10^7$]~M$_{\odot}$ (\citeauthor{Wood:2008hx}\citeyear{Wood:2008hx}), corresponding\footnote{M$_{\rm vir}$ and c$_{\rm vir}$ are strongly correlated (\citeauthor{Navarro:1996gj} \citeyear{Navarro:1996gj}, \citeauthor{Bullock:1999he} \citeyear{Bullock:1999he}). 
In \citeauthor{Bullock:1999he} (\citeyear{Bullock:1999he}) ,
$\rm c_{\rm vir} = 9 \times (M_{\rm vir}/1.5\times 10^{13}\rm h^{-1}M_{\odot})^{-0.13}$ where h is the present day normalized Hubble constant \citep{pdg}.}
to c$_{\rm vir}$ between $\sim$48 and 65. In this paper, initial DM haloes of GCs are modelled with $\rm M_{\rm vir} = 10^7$ M$_{\odot}$. The value of 
$c_{\rm vir}$ used in the model of NGC 6388 is calculated from the formula of \citeauthor{Bullock:1999he} (\citeyear{Bullock:1999he}). 
For M 15, the value of $c_{\rm vir}$ is taken from \citeauthor{Wood:2008hx}(\citeyear{Wood:2008hx}). Both values are listed in  Tab.~\ref{tab:jbarvalue}.

The presence of a central BH changes the DM and stellar densities in regions    
where the BH dominates the gravitational potential, {\it i.e.} for distances to the BH lower than the radius 
of gravitational influence $r_{\rm h}$\footnote{The radius of gravitational influence of a BH is defined by
the equation $\rm M(<r_{\rm h}) \equiv \int_0^{r_h}\rho(r) d^3r = 2\,M_{\rm BH}$.}. 
The adiabatic growth of the BH leads to a spiked DM distribution with an index of 9/4 for an initial 
DM distribution with an index of 1, as for the NFW profile. This profile is referred to as the IMBH NFW profile. The spike is smoothed by the kinetic heating of DM by stars over the timescale $T_{\rm r}$, 
forming a density profile proportional to r$^{\rm -3/2}$ called DM crest (\citeauthor{Merritt:2006mt} \citeyear{Merritt:2006mt}), which corresponds to the final profile. 

The dark halo models of M~15 and NGC~6388 are described in details in the 
appendix. The DM halo of M~15 differs from the model published in \citeauthor{Wood:2008hx} (\citeyear{Wood:2008hx}), since the effect of dark matter heating
by stars is considered in addition to the effect of adiabatic contraction. 
The inferred DM mass densities of  NGC~6388 and M~15 are shown in Fig.~\ref{fig:Density_NGC6388} and Fig.~\ref{fig:Density_M15} respectively.

\subsection{Dark matter annihilation signal}\label{sec:dma}
The gamma-ray flux expected from DM annihilations can be decomposed into an 
astrophysical term and a particle physics term as \citep{Bertone:2004pz}:
\begin{equation}
\label{eqnp}
\frac{\mathrm{d}\Phi(\Delta\Omega,E_{\gamma})}{\mathrm{d}E_{\gamma}}\,=\frac{1}{8\pi}\,\underbrace{\frac{\langle
\sigma
v\rangle}{m^2_{\rm DM}}\,\frac{\mathrm{d}N_{\gamma}}{\mathrm{d}E_{\gamma}}}_{Particle\,
Physics}\,\times\,\underbrace{\bar{J}(\Delta\Omega)\Delta\Omega}_{Astrophysics}\, .
\end{equation}
The astrophysical factor ($\bar{J}$) is generally expressed as the integral over the line of sight ($los$) of 
the squared density averaged over the solid angle $\Delta\Omega$: 
\begin{equation}
\label{eqn:jbar}
\bar{J} = \frac{1}{\Delta\Omega} \int_{\Delta\Omega}d\Omega\int_{los}ds\,\rho^2(r(s)) 
\end{equation}
with
$r(s)~= ~\sqrt{s^2+s_0^2-2ss_0\cos\theta}$, $s_0$ the distance of the source from the Sun and
$\theta$ the opening angle of the integration cone centered on the target position.
To match the analysis cuts used in this work (see Section 2),
$\Delta\Omega$ is set to $\rm5\times10^{-6}\,sr$. 
Tab.~\ref{tab:jbarvalue} shows the values of the 
astrophysical factor  for the DM halo profiles presented in Sec.~\ref{sec:dmm}. 
In the case of the IMBH NFW and final DM profiles for NGC~6388, 
the calculation of the astrophysical factor requires a minimum cutoff for the integration radius.
For the IMBH NFW and final profiles, the integral diverges as $r_{\rm \min}^{-3/2}$ and $\log(r_{\rm \min}^{-1}$) respectively, where  $r_{\rm \min}$
is the inner radius. $r_{\rm \min}$ is usually taken as Max$[r_{S}, r_{A}]$ where $r_S \equiv  2 G M_{\rm BH}/c^2$ is the Schwarzschild radius of the black hole and 
$r_{\rm A}$ is the  self-annihilation radius calculated for an annihilation time of 10 Gyr. Typical values of $m_{DM}$ and $\langle \sigma v\rangle$ give r$_{\rm A}$ $\simeq$ 10$^{-5}$ pc
so that $r_{\rm \min}$ = $r_{\rm A}$.
The value of the astrophysical factor for the final profile is insensitive to the assumed value of $r_{\rm min}$. 
The values for M~15 are calculated using the DM modelling described in Sec.~\ref{sec:dmm}.
The evolution of M~15 leads to a depletion of DM, implying a decrease of $\bar{J}$.
In the case of NGC~6388, the effect of the BH in the stellar environment 
boosts $\bar{J}$ to a value higher than that obtained for the initial NFW profile.

\subsection{Exclusion limits}
The exclusion limits on the particle physics parameters, {\it i.e.} the DM particle mass $m_{\rm DM}$ and the velocity-weighted annihilation cross section
$\langle \sigma v\rangle$, are calculated using the astrophysical factor calculated for each DM halo model by:
\begin{equation}
\langle \sigma v \rangle_{\rm min}^{\rm 95\%\,
C.L.}\,=\,\frac{8\pi}{T_{\mathrm{obs}}}\frac{m^2_{\rm DM}}{\bar{J}(\Delta
\Omega)\Delta\Omega}\,\frac{N_{\gamma}^{\rm 95\%\,C.L.}}{\displaystyle
\int_{0}^{m_{\rm DM}}A_{\mathrm{eff}}(E_{\gamma})\frac{\mathrm{d}N_{\gamma}}{\mathrm{d}E_{\gamma}}dE_{\gamma}}
\end{equation}
where $\mathrm{d}N_{\gamma}/\mathrm{d}E_{\gamma}$ is the DM-produced differential continuum gamma-ray spectrum 
and $A_{\rm eff}$ is the effective area of the instrument during the observations \citep{Aharonian:2007km}.
 Fig.~\ref{fig:ExclusionLimits_NGC6388}
shows the 95\% C.L. exclusion limits for NGC~6388 on $\langle \sigma v\rangle$ for the initial NFW (dashed line) and the final profile
(solid line) derived from the 95\% C.L. upper limit on the number of gamma-rays.
The generic parametrization from \citeauthor{Bergstrom:2000pn} (\citeyear{Bergstrom:2000pn}) as well as a parametrization including 
contributions from virtual internal Bremsstrahlung and final state radiation \citep{Bringmann:2007nk}
are used for the differential gamma-ray spectra. 
The \citeauthor{Bergstrom:2000pn} (\citeyear{Bergstrom:2000pn}) parametrization is derived from a fit to the gamma-ray spectrum 
from WIMP annihilations into  W and Z pairs.
The latter includes both gamma-rays from virtual particles and from charged particle final states of the pair annihilation of winos \citep{Bertone:2004pz}.
The limits are 1 to 3 orders of magnitude above the natural value of the velocity-weighted annihilation cross section for thermally-produced 
DM \citep{Bertone:2004pz}.
Fig.~\ref{fig:ExclusionLimits_M15} shows the H.E.S.S. 95\% C.L. exclusion limits for M~15 for the initial and final DM profiles,
as well as those obtained with the Whipple Cherenkov telescope 
(blue area) in \citeauthor{Wood:2008hx}(\citeyear{Wood:2008hx}).
The thickness of the drawn lines represents the astrophysical uncertainty induced by the plausible mass range for the initial virial mass.  The H.E.S.S. limit 
reaches $\langle\sigma v\rangle$ $\sim$~$\rm 5\times10^{-23}\,cm^{3}s^{-1}$  and $\langle\sigma v\rangle$ $\sim$~$\rm 5\times10^{-24}\,cm^{3}s^{-1}$ around $m_{\rm DM}$ = 2 TeV for the initial NFW profile  and the final profile respectively. For comparison, the exclusion limit obtained for H.E.S.S. using the 
DM halo modelling  of \citeauthor{Wood:2008hx}(\citeyear{Wood:2008hx}) are also shown (gray area). Stronger constraints are obtained with H.E.S.S. due to the combination of larger effective area, $\rm 3\times10^5\,m^2$ vs. $\rm 5\times10^4\,m^2$, better angular resolution, 
0.06$^{\circ}$ vs. 0.25$^{\circ}$, better upper limit on the number of gamma-rays (11.5 vs. $\sim$ 67.7) 
and longer observation time (15.2 h vs. 1.2 h).

\section{\label{sec:summary}Summary}

The present paper gives for the first time exclusion limits on dark matter towards several globular clusters taking into account all relevant astrophysical effects affecting 
 the hypothetical DM halo.
The H.E.S.S. observations reveal no significant gamma-ray excess from point-like sources located at the nominal position of the Galactic GCs NGC~6388 and M~15.
The hypothetical DM halo has been modelled taking into account possible astrophysical processes leading to substantial changes in the initial DM profile: 
the adiabatic contraction of DM by baryons and the adiabatic growth of a BH at the center of the DM halo. 
The scattering of DM by stars in such a dense stellar environment has been taken into account to provide
realistic final DM haloes. This effect is of crucial importance to model DM haloes in these baryon-dominated environments and leads to a depletion of DM during the evolution of the globular cluster. On the other hand, the presence of a central massive BH enhances the DM density in the center.
The constraints on the velocity-weighted annihilation cross section of the DM particle are derived using DM halo profiles taking into account the above-mentioned astrophysical effects on the initial DM density. They lie at the level of a few 10$^{-25}$\,cm$^{3}$s$^{-1}$ for NGC~6388 in the TeV energy range. 
Assuming the absence of a massive BH in the center of M~15, the constraints are of the order of a few 10$^{-24}$\,cm$^{3}$s$^{-1}$.

\acknowledgments
The support of the Namibian authorities and of the University of
Namibia in facilitating the construction and operation of H.E.S.S.
is gratefully acknowledged, as is the support by the German
Ministry for Education and Research (BMBF), the Max Planck
Society, the French Ministry for Research, the CNRS-IN2P3 and the
Astroparticle Interdisciplinary Programme of the CNRS, the U.K.
Particle Physics and Astronomy Research Council (PPARC), the IPNP
of the Charles University, the South African Department of Science
and Technology and National Research Foundation, and by the
University of Namibia. We appreciate the excellent work of the
technical support staff in Berlin, Durham, Hamburg, Heildelberg,
Palaiseau, Paris, Saclay, and in Namibia in the construction and
operation of the equipment.

\appendix

\section{Models of the M~15 and NGC~6388 dark matter halos}
\subsection{M~15 dark matter halo}
The modelling of the M~15 DM halo proceeds in two steps. In the first step, the dark halo is assumed to be adiabatically compressed during the collapse of the core of M~15. The model used for the initial and final baryon and DM densities is described in~\citeauthor{Wood:2008hx} (\citeyear{Wood:2008hx}).
The final baryon density is the observed mass density, taken from~\citeauthor{Gebhardt:1996gp} (\citeyear{Gebhardt:1996gp}).
The DM is compressed during the baryonic collapse. The timescale for the collapse is $\simeq$ 100  $T_{\rm r},$ where the relaxation time $T_{\rm r}$ is given by \citeauthor{spitzer} (\citeyear{spitzer}):
\begin{equation}
T_{\rm r} = \frac{3.4 \times 10^{9}}{ \ln{\Lambda}}    \Big(\frac{v_{\rm rms}}{\rm kms^{-1}}\Big)^3    \Big(\frac{m}{\rm M_{\odot}}\Big)^{-2}       \Big(\frac{n}{\rm pc^{-3}}\Big)^{-1} \mbox{yr} \; .
\label{eq:relaxationtime}
\end{equation} 
In Eq.~(\ref{eq:relaxationtime}), $v_{\rm rms}$ is the velocity dispersion, $n$ is the stellar
density and $\ln{\Lambda}$ is the usual Coulomb logarithm.
In the case of the center of M~15, taking $v_{\rm rms}$ = 10.2$\pm$1.4\,kms$^{-1}$~(\citeauthor{1997ApJ...481..267D}\citeyear{1997ApJ...481..267D}), $\ln{\Lambda}=13.1$ and adopting a typical stellar 
mass value of $m$ = 0.4\,M$_{\odot}$~(\citeauthor{1997ApJ...481..267D}\citeyear{1997ApJ...481..267D}), one finds $T_{\rm r}\simeq\,7\times$10$^{4}$ yr. $T_{\rm r}$ is an increasing function of the distance $r$ to the center of the M~15.   For $r \gtrsim r_{\rm heat}$ = 5\,pc, the relaxation time is larger than the age of the Universe. 
The central value of $T_{\rm r}$ and the position of r$_{\rm heat}$ have only weak dependencies on the actual values of $v_{\rm rms}$ and $n$ when the latter are varied in their uncertainty ranges.
Because of AC, the DM evolution takes place in less than a few orbital periods. 
The orbital period of a star orbiting the core of M~15 is of the order of 1000\,yr, which is much less
than $T_{\rm r}.$ The AC method should thus be valid.  The value of the DM density after AC is similar to the result 
of \citeauthor{Wood:2008hx}(\citeyear{Wood:2008hx}), see Fig.~\ref{fig:Density_M15} and Fig.~7 of \citeauthor{Wood:2008hx}( \citeyear{Wood:2008hx}).
At the same time, the DM is heated up by stellar matter. This process is described in \citeauthor{Merritt:2003qk} (\citeyear{Merritt:2003qk}). 
DM is scattered by stars in a few $T_{\rm r},$ leading to a depletion of the core. 
For $r \gtrsim$ 5\,pc, the DM distribution is not affected by heating. The DM scattering is taken into account with the procedure described in \citeauthor{Bertone:2008a} (\citeyear{Bertone:2008a}) for the M~4 globular cluster.  
A DM mass density of $\rho_{\rm M15}\;\sim$ 35\,$\rm M_{\odot}pc^{-3}$  is obtained at the radius where the heating time is comparable to the age of the universe.
For $r \lesssim$ 5 pc, the DM halo is swept out by heating and the DM mass density was assumed to take the constant $\rho_{\rm M15}$ value.  
The DM mass density of  M~15 called final profile is shown on Fig.~\ref{fig:Density_M15}. 
\subsection{NGC~6388 dark matter halo}
\begin{figure}[]
\epsscale{.8}
\plotone{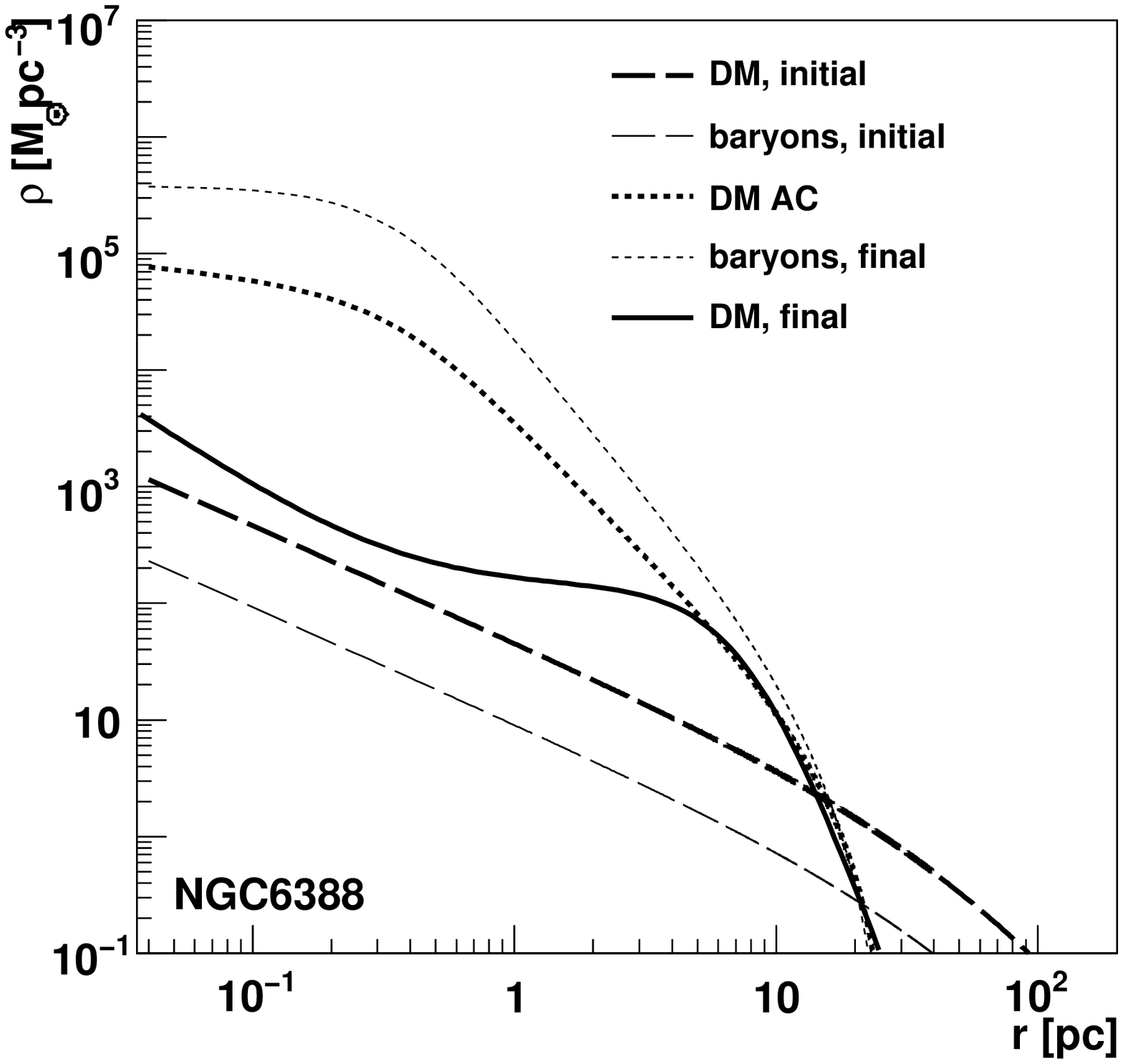}
\caption{DM and baryonic mass density distributions in NGC~6388. 
The DM density before (thick dashed line) and after
(thick dotted line) the adiabatic contraction by baryons is shown. 
The initial DM distribution follows a NFW profile with $\rm M_{\rm vir} = 10^7\,M_{\rm \odot}$. The
initial (thin dashed line) and final  (thin dotted line) baryonic densities are displayed. The final
DM density distribution after the effects of the adiabatic growth of
the IMBH at the center of NGC~6388 and the kinetic heating by stars is presented (thick solid line). See text for more details.
\label{fig:Density_NGC6388}}
\end{figure}

\begin{figure}[t]
\epsscale{.8}
\plotone{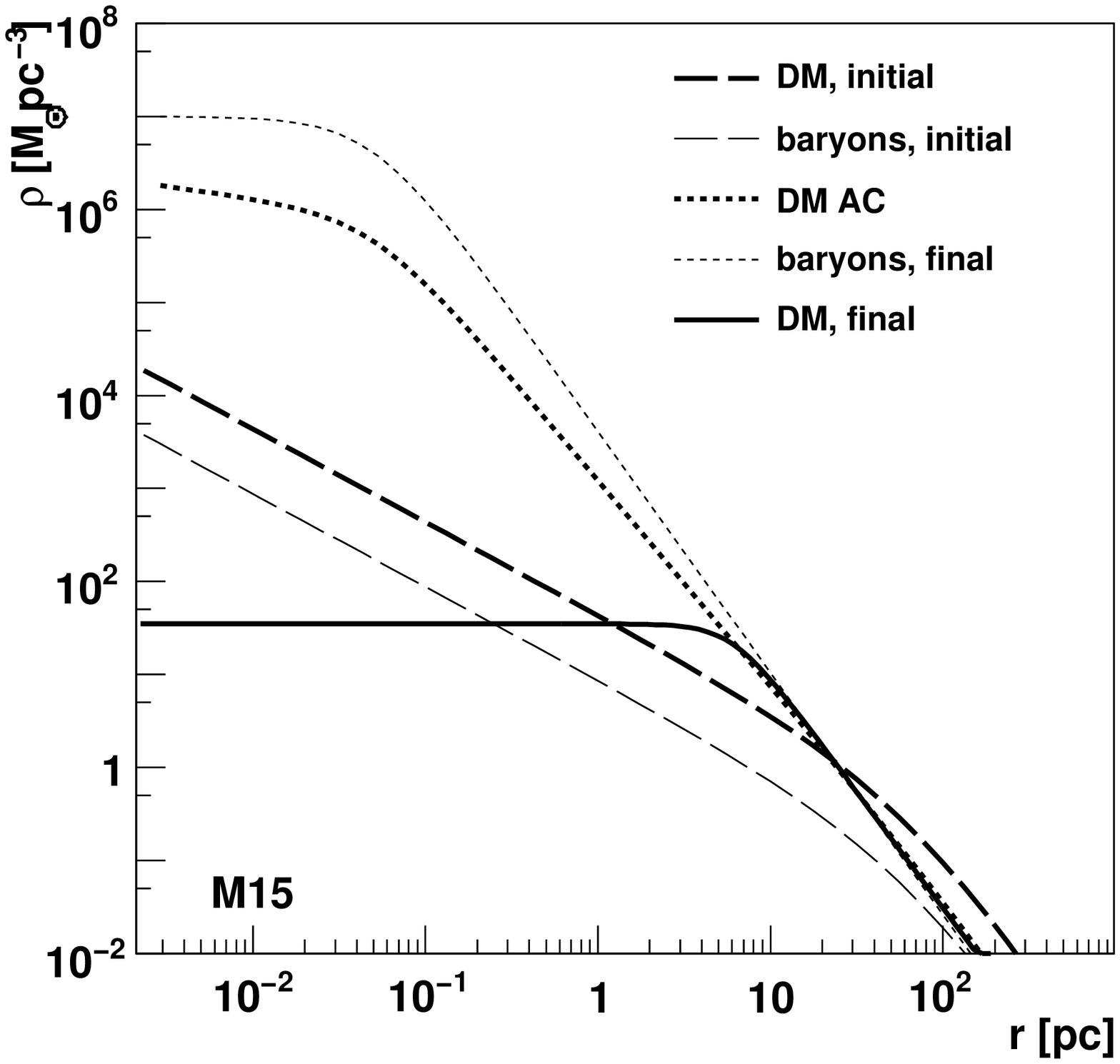}
\caption{DM and baryonic mass density distributions in M~15. 
The DM density before (thick dashed line) and after
(thick dotted line) the adiabatic contraction by baryons is shown. 
The initial DM distribution follows a NFW profile with $\rm M_{\rm vir} = 10^7\,M_{\rm \odot}$. The
initial (thin dashed line) and final  (thin dotted line) baryonic densities are displayed. The final
DM density distribution after the effect of the kinetic heating by stars is presented (thick solid line). See text for more details.
\label{fig:Density_M15}}
\end{figure}

\begin{figure}[]
\epsscale{.8}
\plotone{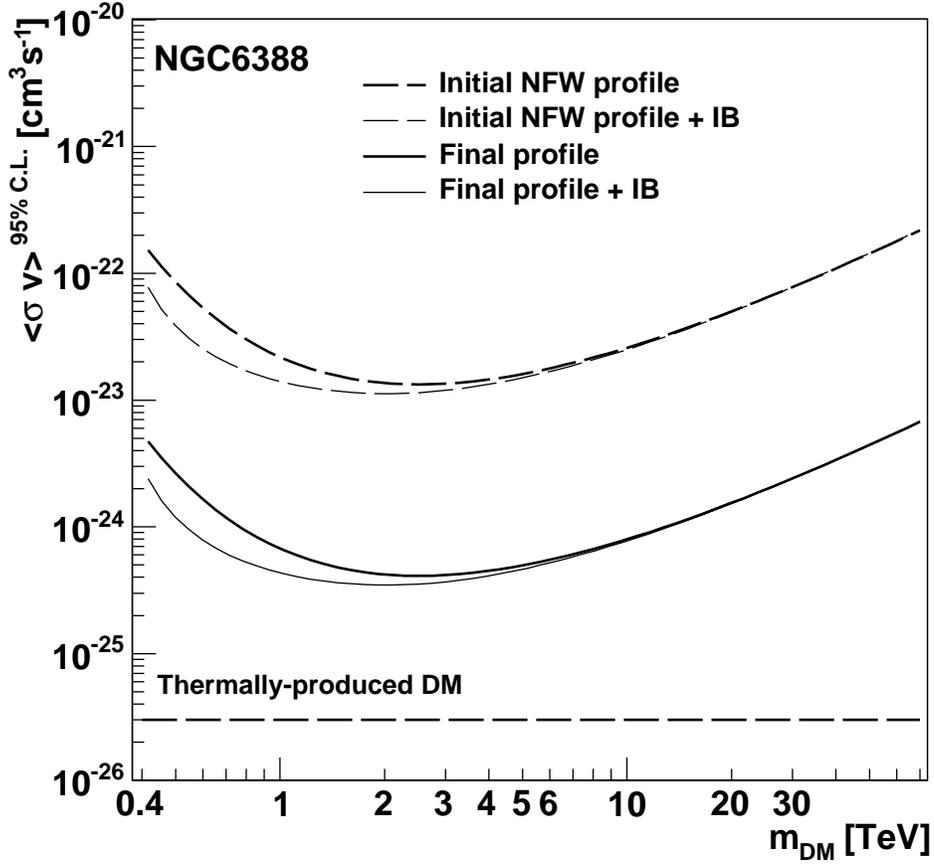}
\caption{H.E.S.S. upper limits at 95\% C.L. on the velocity-weighted annihilation cross section $\langle\sigma v\rangle$ 
versus the DM mass $m_{\rm DM}$ for the Galactic GC NGC~6388. DM halo profiles shown here  
correspond to the initial NFW profile  (dashed thick line) and the realistic profile taking into account plausible astrophysical effects (solid thick line).  
The contribution from internal Bremsstrahlung  and final state radiation to the annihilation spectrum is also shown (dashed/solid thin lines) for both profiles.
The natural value of $\langle\sigma v\rangle$ for thermally-produced DM is also displayed (long-dashed line).\label{fig:ExclusionLimits_NGC6388}}
\end{figure}

\begin{figure}[]
\epsscale{.8}
\plotone{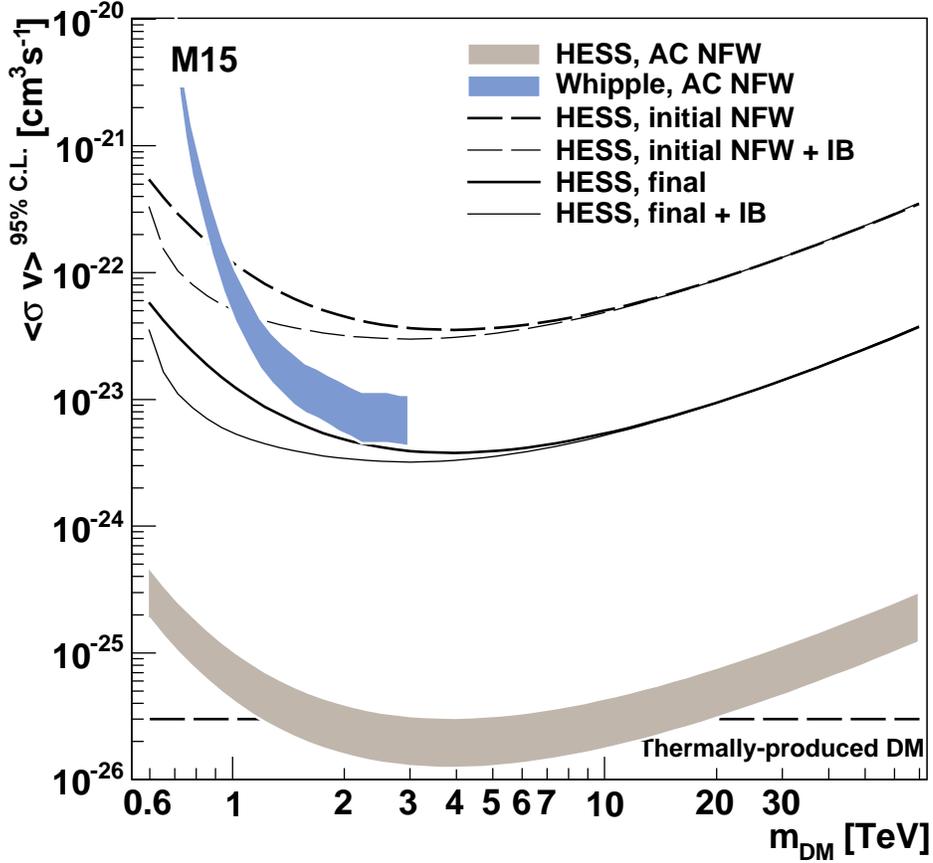}
\caption{H.E.S.S. upper limits at 95\% C.L. on the velocity-weighted anihilation cross section $\langle\sigma v\rangle$ 
versus the DM mass $m_{\rm DM}$ for the Galactic GC M~15. Three DM haloes are shown: 
the initial NFW profile  (dashed thick/thin line), the NFW profile after DM adiabatic contraction by baryons (AC NFW)
used in~\citeauthor{Wood:2008hx} (\citeyear{Wood:2008hx}) (gray area), and the final DM profile (solid thick/thin line). The effect of internal Bremsstrahlung is also presented (thin dashed/solid lines).
 The Whipple exclusion limits extracted from~\citeauthor{Wood:2008hx}  (\citeyear{Wood:2008hx}) is also plotted (blue area).  The natural value of $\langle\sigma v\rangle$ for thermally-produced DM is shown (long-dashed line).\label{fig:ExclusionLimits_M15}}
\end{figure}

The modelling of the NGC~6388 DM halo also proceeds in two steps.
The first step is the AC of the DM halo by the IMBH and baryons. 
An initial baryon fraction of 20\%~\citep{Spergel:2006hy} is assumed with the same spatial distribution like the DM.  The AC scenario gives
the resulting DM distribution knowing the measured baryonic mass profile. This DM halo profile is called AC NFW profile.
The surface density profile of NGC~6388 is well fitted by a modified King model including a black hole, characterized by
a core radius $r_{\rm c}$ = 7.2$^{\prime\prime}$ and a concentration $c$ = 1.8~(\citeauthor{Lanzoni:2007ms}\citeyear{Lanzoni:2007ms}). Using these parameters, the numerical integration
of the Poisson equation yields the behavior of the gravitational potential from which it is straightforward to compute the baryonic density profile.
 The initial NFW profile is characterized by $\rm M_{\rm vir} = 10^7\,M_{\rm \odot}$.
 Since the mass of the IMBH is just a small fraction of the total mass in the core of NGC~6388, the dynamics of DM is influenced mainly by baryonic matter,
except in the immediate vicinity of the black hole. Using a central velocity dispersion of $v_{\rm rms}$ = 18.9$\pm$0.8 kms$^{-1}$ (\citeauthor{Pryor:1993}, \citeyear{Pryor:1993})
and  $\ln{\Lambda}=14.7$, the central relaxation time is found to be $T_{\rm r}$
$\simeq$ 8$\times$10$^{6}$ yr. The orbital period of a star orbiting the core of NGC~6388 is 5000\,yr, so that the AC method is again valid. The relaxation time is larger than the age of the Universe for $r > r_{\rm heat}$. The distribution of the DM density around the black hole (for r $\lesssim$ r$_{\rm h}$) is changing with time, but tends to a power law with index 3/2 after a few $T_{\rm r}.$ The final DM distribution is thus obtained by extending the prescription of  \citeauthor{Bertone:2008a} (\citeyear{Bertone:2008a}).  Far from the center of the cluster, the stellar density is low and thus the 
heating time becomes large so that the DM distribution is unaffected. A mass density of $\sim$140~$\rm M_{\odot}pc^{-3}$ is obtained at the radius $r_{\rm heat}$ $\sim$ 4\,pc  
where the heating time is comparable to the age of the Universe. In the region between $r_{\rm h} < r <  r_{\rm heat}$, the DM density 
is expected to be described by a smooth curve similar to Fig.~1 of \citeauthor{Merritt:2006mt}(\citeyear{Merritt:2006mt}). 
In the modelling for NGC~6388, the DM density was conservatively assumed to take a
constant value of 140\,$\rm M_{\odot}pc^{-3}$ in the region $r_{\rm h} < r < r_{\rm heat}$. 
For  $r < r_{\rm h}$, the DM density is given by  $\rm  \rho(r)= 140\,M_{\odot}pc^{-3} \,(r/r_{\rm h})^{-3/2}$.
The final profile for the DM distribution of NGC~6388 is shown in Fig.~\ref{fig:Density_NGC6388}.
At the position of NGC~6388, the DM density from the smooth Galactic halo assuming a NFW profile is $\sim$0.03\,$\rm M_{\odot}pc^{-3}$.

\begin{table}[!hb]
\begin{center}
\caption{List of H.E.S.S. analysis cuts~\citep{deNaurois:2009ud}. The shower depth is the reconstructed primary interaction depth of the particle and the nominal distance is the angular distance
of the image barycenter to the camera center.\label{tabcut}}
\begin{tabular}{cc}
\\
\tableline\tableline
 Cut name& $\gamma$-event cut value \\
 \tableline\tableline
Shower goodness&$\le$ 0.4\\
Image charge (photo-electrons)& $\ge$ 120\\
Reconstructed shower depth (radiation length)& $\rm[-1,4]$\\
Reconstructed nominal distance ($^{\circ}$)& $\le$ 2\\
Reconstructed event telescope multiplicity & $\ge$ 2\\
\tableline
\tableline
\end{tabular}
\end{center}
\end{table}

\begin{table}[hb]
\begin{center}
\caption{Properties of the NGC~6388 and M~15 globular clusters used in this study.\label{tab0}}
\begin{tabular}{lcc}
\\
\tableline\tableline
& NGC~6388& M~15 \\
 \tableline\tableline
 Other Name &/& NGC~7078\\
RA, Dec coordinates (J2000)&$\rm17^h36^m17.05^s$, $\rm -44^{\circ}44'05.8''$&$\rm 21^h28^m58.3^s$, $\rm 12^{\circ}10'00.6''$\\
Galactic coordinates (l, b)&345.55$^{\circ}$, -6.73$^{\circ}$&64.8$^{\circ}$, -27.1$^{\circ}$\\
Distance (kpc) &11.5\tablenotemark{a}&10.0\tablenotemark{c}\\
Core radius r$_{\rm c}$ (pc) &0.4\tablenotemark{a}&0.04\tablenotemark{d}\\
Tidal radius r$_{\rm t}$ (pc) &25\tablenotemark{a}&30\tablenotemark{c}\\
Estimated mass (M$_{\odot}$) &10$^6$&5$\times$10$^5$\\
Core stellar density (M$_{\odot}$pc$^{-3}$) &5$\times$10$^5$ &10$^7$\\
Central velocity dispersion v$_{\rm rms}$ (kms$^{-1}$) & 18.9$\pm$0.8\tablenotemark{b}&10.2$\pm$1.4\tablenotemark{d}\\
\tableline
\tableline
\end{tabular}
\tablenotetext{a}{\citeauthor{Lanzoni:2007ms}(\citeyear{Lanzoni:2007ms})}
\tablenotetext{b}{\citeauthor{Pryor:1993} (\citeyear{Pryor:1993})}\tablenotetext{c}{\citeauthor{Wood:2008hx}(\citeyear{Wood:2008hx})}\tablenotetext{d}{\citeauthor{1997ApJ...481..267D}(\citeyear{1997ApJ...481..267D})}
\end{center}
\end{table}

\begin{table}[hb]
\begin{center}
\caption{Exposure time, mean zenith observation angle,  number of ON events, number of OFF events, $\alpha$ ratio and 95\% C.L. upper limits on the number of gamma-rays from the H.E.S.S. observations 
for NGC~6388 and M~15, respectively.  See text for more details.\label{tab1}}
\begin{tabular}{lcc}
\\
\tableline\tableline
 & NGC~6388& M~15 \\
 \tableline\tableline
Exposure time\tablenotemark{a} (hours)&27.2&15.2\\
Mean zenith observation angle &22.9$^{\circ}$ &37.0$^{\circ}$\\
N$_{\rm ON}$& 75& 28 \\
N$_{\rm OFF}$& 1754 & 719 \\
$\alpha$& 26.5 & 26.6 \\
$\rm N_{\gamma}^{95\% C.L.}$& 21.6 & 11.5 \\
\tableline
\tableline
\end{tabular}
\tablenotetext{a}{After quality selection.}
\end{center}
\end{table}

\begin{table}[!ht]
\begin{center}
\caption{Values of the los-integrated squared density averaged over the solid angle ($\bar{J}$) 
expressed in units of 10$^{24}$ GeV$^2$cm$^{-5}$, for the different DM halo profiles. The integration solid angle is 
$\rm \Delta\Omega = 5\times10^{-6}$ sr. The virial mass and concentration for the initial NFW profiles are given in brackets. \label{tab:jbarvalue}}
\begin{tabular}{ccc}
\\
\tableline\tableline
DM halo profile name & NGC~6388 & M~15\\ 
\tableline
Initial NFW (M$_{\rm vir}$, c$_{\rm vir}$)& 2.1 (10$^7$ M$_{\odot}$, 60)& 1.5 (10$^7$ M$_{\odot}$, 50)\\
AC NFW&1.3$\times$10$^4$& 4.3$\times$10$^3$\\
IMBH NFW&2.2$\times$10$^4$&/\\
Final&68&14\\
\tableline\tableline
\end{tabular}
\end{center}
\end{table}

\begin{thebibliography}{1}

\bibitem[Aharonian et al. 2005]{Aharonian:2004ap} Aharonian, F., {\it al.}  [HESS Collaboration], 2005, \aap, 430, 865 
\bibitem[Aharonian et al. 2006a]{Aharonian:2006wh} Aharonian, F., {\it et al.}  [HESS Collaboration], 2006a, \prl, 97, 221102 [Erratum-ibid.\  97, 249901]
\bibitem[Aharonian et al. 2006b]{Aharonian:2006pe} Aharonian, F., {\it et al.}  [HESS Collaboration], 2006b,  \aap, 457, 899
\bibitem[Aharonian et al. 2008]{Aharonian:2007km} Aharonian, F., {\it et al.}  [HESS Collaboration], 2008,  Astropart.\ Phys., 29, 55 [Erratum-ibid.\  2010, 33, 174]
\bibitem[Aharonian et al. 2009a]{Aharonian:2009zk} Aharonian, F.,  [H.E.S.S.collaboration], 2009a, \aap,  503, 817
\bibitem[Aharonian et al. 2009b]{Aharonian:2008dm} Aharonian, F., {\it et al.}  [HESS Collaboration], 2009b, \apj, 691, 175
\bibitem[Aharonian et al. 2011]{Aharonian:2010zzt} Aharonian, F., {\it et al.}  [HESS Collaboration], 2011, Astropart.\ Phys., 34, 608 
\bibitem[Aharonian et al. 2009b]{Aharonian:2009nh} Aharonian, F., {\it et al}  [HESS collaboration], 2009b, \aap, 499, 273 
\bibitem[Amsler et al. 2009]{pdg} Amsler, C., {\it et al.} (Particle Data Group), 2008, Phys. Lett. B, 667, 1
\bibitem[Anderhub et al. 2009]{Anderhub:2009px} Anderhub, H., {\it et al.}  [MAGIC Collaboration], 2009, \apj, 702, 266
\bibitem[Baumgardt et al. 2005]{Baumgardt:2004tp} Baumgardt, H., Makino, J., \& Hut, P., 2005, \apj, 620, 238
\bibitem[{{Baumgardt} {et al.}(2009)}]{Baumgardt:2009hv} {Baumgardt}, H., et al. 2009, \mnras, 396, 2051 
\bibitem[Berge et al. 2007]{Berge:2006ae} Berge, D., Funk, S., \& Hinton, J., 2007, \aap, 466, 1219
\bibitem[Bergstr{\"o}m(2000)]{Bergstrom:2000pn} Bergstr{\"o}m, L., 2000, Reports on Progress in Physics, 63, 793 
\bibitem[{{Bertone \& Fairbarn}(2008)}]{Bertone:2008a} Bertone, G., \& Fairbarn, M.,  2008, \prd, 77, 043515
\bibitem[Bertone et al. 2005]{Bertone:2004pz} Bertone, G., Hooper, D., \&  Silk, J., 2005, Phys.\ Rept.\  405, 279 
\bibitem[Binney \& Tremaine 1987]{binneytremaine} Binney, J., \& Tremaine, S.,  1987, Princeton, NJ, Princeton University Press, 1987, 747 p.
\bibitem[Blumenthal et al. 1986]{Blumenthal:1985qy} Blumenthal, G. R., Faber, S. M., Flores, R., \& Primack, J. R., 1986, \apj, 301, 27 
\bibitem[Bringmann et al. 2008]{Bringmann:2007nk} Bringmann, T., Bergstrom, L., \& Edsjo, J., 2008, JHEP, 0801, 049
\bibitem[Brodie \& Strader 2006]{Brodie:2006zz} Brodie, J.~P. \& Strader, J., 2006, ARA \& A, 44, 193
\bibitem[{{Bullock} {et al.}(2001)}]{Bullock:1999he} Bullock, J. S., {\it et al.}, 2001, \mnras, 321, 559 
\bibitem[{{Cohen} {et al.}(2010)}]{Cohen:2010gq} Cohen, J.~G., Kirby, E.~N., Simon,  J.~D., \& Geha, M., 2010, \apj, 725, 288 
\bibitem[Colbert \& Ptak 2002]{Colbert:2002mi} Colbert, E., \& Ptak, A., 2002, \apjs,  143, 25
\bibitem[{{Conroy} {et al.}(2010)}]{Conroy:2010bs} Conroy, C., Loeb, A. \& Spergel, D., 2010, arXiv:1010.5783
\bibitem[De Paolis et al. 1996]{DePaolis:1996tq} De Paolis, F., Gurzadian, V. G., \& Ingrosso, G., 1996, \aap, 315, 396
\bibitem[Dewangan et al. 2005]{Dewangan:2005kj} Dewangan, G. C., {\it et al.}, 2005, \apj, 635, 198 
\bibitem[{{Dull} {et al.}(1997)}]{1997ApJ...481..267D} Dull, J.~D., {\it et al.}, 1997, \apj, 481, 267 
\bibitem[Englmaier \& Gerhard 2005]{Englmaier:2006pa} Englmaier, P., \& Gerhard, O.,  2005,  American Astronomical Society, 36$^{\rm th}$ Annual Meeting, Santa Barbara, California,
arXiv:astro-ph/0601679
\bibitem[{{Gebhardt} {et al.}(1997)}]{Gebhardt:1996gp} Gebhardt, K., Pryor, C., Williams, T. B., Hesser, J. E., \& Stetson, P. B., 1997, \aj, 113, 1026 
\bibitem[Gerssen et al. 2002]{Gerssen:2002iq} Gerssen, J., {\it et al.}, 2002, \aj, 124, 3270
\bibitem[Gnedin et al. 2004]{Gnedin:2004cx} Gnedin, O. Y., Kravtsov, A. V., Klypin, A. A., \& Nagai, D., 2004, \apj, 616, 16
\bibitem[Gnedin \& Primack 2004]{Gnedin:2003rj} Gnedin, O. Y., \& Primack, J. R., 2004, \prl, 93, 061302
\bibitem[Gondolo \& Silk 1999]{Gondolo:1999ef} Gondolo, P., \& Silk, J., 1999, \prl, 83, 1719
\bibitem[Harris 1996]{Harris:1996} Harris, W. E. 1996, \aj, 112, 1487 (Dec. 2010 update) 
\bibitem[Islama et al. 2003]{islama:2003} Islama, R., Taylor, J., \& Silk, J., 2003, \mnras, 340, 6471
\bibitem[Jesseit et al. 2002]{Jesseit:2002tj} Jesseit R., Naab, T., \& Burkert, A., 2002, \apj, 571, L89
\bibitem[Kabuki et al. 2007]{Kabuki:2007am} Kabuki, S., {\it et al.} [CANGAROO Collaboration], 2007, \apj, 668, 968
\bibitem[Kiselev et al. 2008]{Kiselev2008} Kiselev, A.~A., Gnedin, Y.~N., Shakht, {\it et al.}, 2008, Astronom. Lett., 34, 529 
\bibitem[Komatsu et al. 2009]{Komatsu:2008hk} Komatsu, E., {\it et al.}  [WMAP Collaboration], 2009, \apjs, 180, 330
\bibitem[Kong 2007]{Kong:2007mu} Kong, A. K. H., 2007, \apj, 668, L139
\bibitem[Koushiappas et al. 2004]{Koushiappas:2003zn} Koushiappas, S. M., Bullock, J. S., \& Dekel, A., 2004, \mnras, 354, 292
\bibitem[{{Lanzoni} {et al.}(2007)}]{Lanzoni:2007ms} Lanzoni, B., {\it et al.}, 2007, \apj, 668, L139
\bibitem[McCutcheon 2009]{McCutcheon:2009hs} McCutcheon, M. [VERITAS Collaboration], Proc. of the 31$^{st}$ ICRC, Lodz, Poland, July 2009, arXiv:0907.4974 [astro-ph.HE]
\bibitem[Magorrian et al. 1998]{Magorrian:1997hw} Magorrian, J., {\it et al.}, 1998,  \aj, 115, 2285
\bibitem[Mateo 1998]{Mateo:1998wg} Mateo, M., 1998, Ann.\ Rev.\ Astron.\ Astrophys., 36, 435
\bibitem[{{Merritt}(2004)}]{Merritt:2003qk} Merritt, D., 2004, \prl, 92, 201304
\bibitem[{{Merritt} {et al.}(2007)}]{Merritt:2006mt} Merritt, D., Harfst, S., \& Bertone, G.,  2007, \prd, 75, 043517
\bibitem[de Naurois \& Rolland 2009]{deNaurois:2009ud} de Naurois, M., \& Rolland, L., 2009, Astropart. Phys., 32, 231
\bibitem[{{Navarro} {et al.}(1997)}]{Navarro:1996gj} {Navarro}, J.~F., {Frenk}, C.~S., \& {White}, S.~D.~M., 1997  \apj, 490, 493 
\bibitem[Noyola \& Gebhardt 2006]{Noyola:2006mg} Noyola, E., \& Gebhardt, K., 2006, \aj, 132, 447
\bibitem[Noyola et al. 2008]{Noyola:2008kt} Noyola, E., Gebhardt, K., \& Bergmann, M., 2008, \apj, 676, 1008 
\bibitem[{{Nucita} {et al.}(2007)}]{Nucita:2007yn} Nucita, A. A., De Paolis, F., Ingrosso, G., Carpano, S., \& Guainazzi, M., 2007, \aap, 763, 478 
\bibitem[Peebles 1984]{Peebles:1984zz} Peebles, P. J. E., 1984, \apj, 277, 470
\bibitem[Prada et al. 2004]{Prada:2004pi}  Prada, F., Klypin, A., Flix, J., Martinez, M., \& Simonneau, E., 2004, \prl, 93, 241301
\bibitem[{{Pryor \& Meylan} (1993)}]{Pryor:1993} Pryor, C \& Meylan, G., 1993, Astronomy Society of the Pacific, 50, 357
\bibitem[P\"uhlhofer et al. 2003]{Puhlhofer:2003ir} P\"uhlhofer, G., {\it et al.}  [HEGRA Collaboration], 2003, Astropart.\ Phys., 20, 267
\bibitem[{{Rolke} {et al.}(2005)}]{rol05} Rolke, W.~A., Lopez, A.~M., \& Conrad, J.,  2005, Nucl. Instrum. Meth., A551, 493 
\bibitem[{{Rosenberg} {et~al.}(2006)}]{rosenberg2006} {Rosenberg}, J.~L., {Ashby}, M.~L.~N., {Salzer}, J.~J., and {Huang}, J.-S., 2006, \apj, 636, 742
\bibitem[Spergel et al. 2007]{Spergel:2006hy} Spergel, D. N., {\it et al.}  [WMAP Collaboration], 2007, \apjs, 170, 377
\bibitem[{{Spitzer}(1987)}]{spitzer} Spitzer, L., 1987, Princeton, NJ, Princeton University Press, 191 p.  
\bibitem[Swartz et al. 2004]{Swartz:2004xt} Swartz, D. A., Ghosh, K. K., Tennant, A. F,. \& Wu, K. W., 2004, \apjs, 154, 519
\bibitem[Ulvestad et al. 2007]{Ulvestad:2007cw} Ulvestad, J. S., Greene, J. E., \& Ho, L. C., 2007, \apj, 661, L151
\bibitem[Umbreit et al. 2008]{Umbreit:2008nm} Umbreit, S., Fregeau, J. M., and Rasio, F. A., 2008, Proc. of IAUS 246 
\bibitem[Volonteri et al. 2003]{volonteri:2003} Volonteri, M., Haardt, F., \& Madau, P., 2003, \apj, 582, 559
\bibitem[{{Wood} {et al.}(2008)}]{Wood:2008hx} Wood, M., {\it et al.}, 2008, \apj, 678, 594 
\bibitem[Yungelson et al. 2008]{Yungelson:2007qu} Yungelson, L. R., van den Heuvel, E. P. J., Vink, J. S., Portegies Zwart, S. F., \& de Koter, A., 2008, \aap, 477, 223
\bibitem[Zeldovich et al. 1980]{Zeldovich:1980st}  Zeldovich, Y. B., Klypin, A. A., Khlopov, M. Y., \& Chechetkin, V. M., 1980,  Sov.\ J.\ Nucl.\ Phys.,  31, 664   [Yad.\ Fiz.,  31, 1286].


\end{thebibliography}
\end{document}